\documentclass[12pt,preprint]{aastex}

\def\({\left(}
\def\){\right)}

\usepackage[usenames]{color}
\usepackage{psfig}
 
\shorttitle{Detection of new point sources in WMAP 7 year data}
\shortauthors{Scodeller, Hansen, Marinucci}

\begin{document}

\title{Detection of new point sources in WMAP 7 year data using internal templates and needlets}

\author{Sandro Scodeller}
\affil{Institute of Theoretical Astrophysics, University of Oslo,
P.O. Box 1029 Blindern, N-0315 Oslo, Norway}
\email{sandro.scodeller@astro.uio.no}

\author{Frode K. Hansen}
\affil{Institute of Theoretical Astrophysics, University of Oslo,
P.O. Box 1029 Blindern, N-0315 Oslo, Norway; \\Centre of Mathematics
for Applications, University of Oslo, P.O. Box 1053 Blindern, N-0316
Oslo }
\email{frodekh@astro.uio.no}

\author{Domenico Marinucci}
\affil{Dipartimento di Matematica, Universit\`a di Roma ``Tor Vergata'', Via
 della Ricerca Scientifica 1, I-00133 Roma, Italy}
\email{marinucc@mat.uniroma2.it}

\begin{abstract}
We have developed a new needlet based method to detect point sources in cosmic microwave background (CMB) maps and have applied it to the WMAP 7 year data. We use both the individual frequency channels as well as internal templates, the difference between pairs of frequency channels, with the advantage that the CMB component is eliminated. Using the area of the sky outside the Kq85 galactic mask, we detect a total of 2102 point sources at the $5\sigma$ level in either the frequency maps or the internal templates. Of these, 1116 are detected either at $5\sigma$ directly in the frequency channels or at $5\sigma$ in the internal templates and $\geq3\sigma$ at the corresponding position in the frequency channels. Of the 1116 sources, 603 are detections which have not been reported so far in WMAP data. We have made a catalogue of these sources available with position and flux estimated in the WMAP channels where they are seen. In total, we identified 1029 of the 1116 sources with counterparts at 5GHz and 69 at other frequencies.
\end{abstract}


\section{Introduction}

The Wilkinson Microwave Anisotropy Probe (WMAP) \citep{WMAP03} measured the Cosmic Microwave Background (CMB) fluctuations at high resolution and signal-to-noise in five frequency bands. The detailed study of the CMB and its anisotropies gives information enabling us to comprehend the universe we live in and how it came to be. It is therefore very important to get the maximum accuracy of the data we have at present. For the moment, the best publicly available data of the CMB is the seven-year WMAP-data \citep{effarea}, but the ongoing Planck mission is expected to improve the quality even more. The data is contaminated by foregrounds, on larger scales the contaminating foregrounds are mainly diffuse galactic emissions, while on smaller scales the main contaminants are extra-galactic point-sources (see for instance \cite{toff98}, \cite{DeZott99}, \cite{Hobs99}, \cite{DeZott05}). Clearly the WMAP-mission in addition to measure the CMB and its anisotropies, provides an all-sky, high frequency survey of diffuse galactic foregrounds and extra-galactic sources. Independently of whether one is interested in studying the  CMB-anisotropy, diffuse galactic emissions or extra-galactic point source measurements, it is crucial to be able to separate the different components. In this paper we are mainly interested in disentangling extra-galactic point sources from the WMAP-data. This has been done by the WMAP team and other teams, here we present a new approach.

\noindent
In \cite{WMAPsrc} the WMAP-collaboration presents two point source catalogues of detected sources obtained from the 7-year data. WMAP finds a total of 542 distinct sources, most of which also have a 5GHz counterpart. They used two different methods: a global filtering method in the five bands (where they find 471 sources) and a CMB-free method (ILC based, introduced and previously applied by \cite{Chen08} to 1-year and 3-year data and applied to 5-year data by \cite{Chen09} for the Q-, V- and W-band (where 417 sources are found); for more details on these approaches we refer to \cite{WMAPsrc} and references therein. Subsequently other approaches have been used by different teams on real data or simulations. The most relevant one to our approach is the one where Mexican wavelets are used to detect sources, introduced by \cite{cayon}; this approach has then been developed further to the Mexican Hat Wavelet Family \cite{Mhwf}. Subsequently, \cite{lopez} successfully applied the Mexican Hat Wavelet (MHW) technique to make a non-blind search in 3 year WMAP-data, where at three sigma level they detect 381 sources at the five-sigma level (98 of which were not present in the WMAP-3year-catalogue). \cite{mass} then apply the MHW to WMAP 5 year data, obtaining 516 point sources at the 5-sigma level.

Other approaches which have been applied include: via Cross-Correlation (\cite{Nie}); via matched filters (\cite{vikh}, \cite{Tegma}, \cite{Barre}, \cite{lopez06}) and matched multifilters (\cite{herr02}, \cite{Lanz11}); Bayesian techniques with prior-information about source distribution (\cite {hobs03}, \cite{Carvalho}, \cite{Arg11}); see also \cite{schmitt2010}, \cite{starck2010} and the references therein for general results on wavelet-based methods to search for point sources in astrophysical data.

Here we present a method for point source detection which is novel in two ways, (1) it uses the family of standard (\cite{need}) and mexican (\cite{scodeller}) needlets optimized for point source detection on the given channels and (2) we search for point sources both in the individual WMAP channels as well as in the CMB-free internal templates constructed from the difference between two frequency channels. 

The outline of the paper is as follows: in section \ref{sec:method} we present in detail how our method for detecting point sources works, in section \ref{sec:simul} we present our results on simulations, and in \ref{sec:wmap} for the real data. Eventually in section \ref{sec:concl} we summarize. In table \ref{tab:new-PS} we list the 1116 sources which we detected in the 5 WMAP channels.

\section{Method}
\label{sec:method}

\subsection{A short introduction to needlets}

Needlets were introduced in the mathematical literature by \cite{npw1} and have recently become a very popular tool for a wide range of CMB analysis tasks, as proved from the variety of statistical procedures where they have been exploited. A partial list includes testing for non-Gaussianity, estimating the angular power spectrum, testing for asymmetries, testing for cross-correlation among CMB and large-scale structure data, map-reconstruction, testing for Bubble Universes; see, for instance \cite{pietro06}, \cite{Baldi09a}, \cite{Baldi09b}, \cite{need}, \cite{Fay08}, \cite{oyst1,oyst2}, \cite{Cabella09}, \cite{Peiris} and \cite{Basak12}. We review briefly their construction, as follows.

Let $b(t)$ be a weight function satisfying three conditions, namely

\begin{itemize}
\item \emph{Compact support}: $b(t)$ is strictly larger than zero only for $%
t\in \lbrack B^{-1},B],$ some $B>1$

\item \emph{Smoothness: }$b(t)$ is $C^{\infty }$

\item \emph{Partition of unity:} for all $l=1,2,...$ we have%
\[
\sum_{j=0}^{\infty }b^{2}(\frac{l}{B^{j}})=1\mbox{ .}
\]
\end{itemize}

Recipes to construct a function $b(t)$ that satisfy these conditions are
easy to find and are provided for instance by \cite{need} and \cite{mape}.
Consider now a grid of points $\left\{ \xi _{jk}\right\} $ on
the sphere and a grid of weights $\lambda _{jk}$; in practice, the points can be viewed as the pixel centres for HEALPix, while the weights can be taken to be constant and equal to the pixel area. The needlet system is then defined by
\[
\psi _{jk}(x)=\sqrt{\lambda _{jk}}\sum_{l=B^{j-1}}^{B^{j+1}}\sum_{m=-l}^{l}b(%
\frac{l}{B^{j}})Y_{lm}(x)\overline{Y}_{lm}(\xi _{jk})\mbox{ ,%
}
\]%
with the corresponding needlet coefficients provided by%
\begin{equation}
\beta _{jk}=\int_{S^{2}}f(x)\psi _{jk}(x)dx=\sqrt{\lambda _{jk}}%
\sum_{l=B^{j-1}}^{B^{j+1}}\sum_{m=-l}^{l}b(\frac{l}{B^{j}}%
)a_{lm}Y_{lm}(\xi _{jk})\mbox{ .}  \label{dirtra}
\end{equation}%

The main features of needlets have now been widely discussed in the literature; 
here, we simply recall the \emph{reconstruction property}
(see \cite{npw1}), entailing that:%
\[
f(x)=\sum_{jk}\beta _{jk}\psi _{jk}(x)\mbox{ .}
\]

More recently, the needlet idea has been extended by \cite{gm1,gm2}, introducing so-called Mexican needlets; loosely speaking, the idea is to replace the compactly supported kernel $b(\frac{l}{B^j})$ by a smooth function of the form
\[
b(\frac{l}{B^j})=(\frac{l}{B^j})^{2p}exp\left({-\frac{l^2}{B^{2j}}}\right) \mbox{ ,}
\]
for some integer parameter $p$, see \cite{scodeller} for numerical analysis and implementation in a  cosmological framework.  Mexican needlets have extremely good localization properties in real space, and for $p=1$ they provide at high frequencies a good approximation to the so-called Spherical Mexican Hat Wavelet construction.

\subsection{Choosing the needlet bases}
In order to amplify the point source signal we use the needlet transform on the maps and search for needlet coefficients with a value larger than 5 times the standard deviation expected from CMB and noise in a given channel or template. As mentioned in the introduction, we will not only look for point sources in the individual WMAP channels, but also in internal templates. The internal template between channel $c$ and channel $c'$, assuming that $c'$ has a smaller beam than $c$, is constructed by smoothing the $c'$ map by the beam $b_\ell^ c/b_\ell^{c'}$. In this way both channels have the same beam, and hence by constructing the difference map between the two, the CMB component disappears. We are thus left with an internal template containing only noise and foregrounds/point sources. The advantages/disadvantages with the two approaches are
\begin{itemize}
\item {\bf individual channels:} The background consist of both CMB and noise. The CMB is dominated by large scale fluctuations and therefore a needlet basis with small extension (high value of $j$) on the sphere is necessary in order to separate the point sources from local CMB fluctuations. Such needlet coefficients of a point source will therefore have a 5 sigma deviation only on a small number of pixels, since the point source in this case will be very localized in needlet space.
\item {\bf internal templates:} The background consists of only noise, but the noise level is higher than in the individual channels, since noise from both channels are present in the template. The absence of dominant large scale fluctuations makes needlets with a larger extension (lower $j$) more efficient. The advantage is that more needlet coefficients will be at 5 sigma for a given source and the probability of a detection is therefore larger. Whereas for the individual channels a point source must have a large amplitude in at least one channel in order to be detected, for the internal template it suffices that the difference in amplitude between the two channels is large. This also implies that we cannot estimate the source amplitude in the template, but we can use the position of a source found in the template to estimate the amplitude in the channel.
\end{itemize}
For each channel and each internal template, we have calculated the needlet coefficients of a simulated point source as well as the standard deviation of needlet coefficients due to CMB and noise (at the given frequency). In that way we are able to calculate the signal-to-noise ratio for a large set of different needlets and find the needlet with the optimal signal-to-noise ratio for point sources for a given channel. In table \ref{tab:optchan} we show which needlets we found optimal for a given channel and template. The templates presented in this table are the templates for which we found the highest signal-to-noise ratio. In addition we will also use the K-Ka template which, even though it does not contain the highest signal-to-noise ratio, is expected to reveal many sources being in the synchrotron dominated frequency range. In the table we also show the distance of influence (for details, see \cite{scodeller}, we used a threshold of $3\%$) which is a measure of angular extension of the needlet on the sphere and is in particular an indication of how extended a point source will appear in given needlet basis.

\begin{deluxetable}{lcc}
\tabletypesize{\normalsize}%\tiny}%scriptsize}%\footnotesize}
\tablecolumns{3}
\tablecaption{Needlet with best signal to noise ratios for channels, needlet for templates with highest signal to noise ratio and distance of influence \label{tab:optchan}}
\tablehead{   % column headings
  \colhead{Needlet}   &
  \colhead{Used in channel/template} &
  \colhead{Distance of influence [deg]\tablenotemark{a}}
}

\startdata
 Mex B=1.9 j=9 &  K       & 1.12  \\
Std B=1.8 j=10 &  Ka      & 1.24  \\
Std B=1.6 j=13 &  Q       & 1.19  \\
 Std B=2 j=9   &  V       & 0.97  \\
Mex B=1.8 j=11 &  W       & 0.86  \\
Mex B=1.9 j=8  &K-Ka,K-V,K-W & 1.59  \\
Mex B=1.8 j=9  & Ka-V      & 1.44  \\
Mex B=2 j=8    & Q-V       & 1.25  \\
\enddata
\tablecomments{Std stands for standard needlets and Mex stands for mexican needlets with $p=1$.}
\tablenotetext{a}{Note that here as opposed to (\cite{scodeller}) the distance of influence is from the center of the source. }
\end{deluxetable}

\subsection{The detection algorithm}
Given the needlet coefficients of a channel or a template at $N_\mathrm{side}=512$, we use the following procedure to detect point sources and estimate amplitudes:
\begin{enumerate}
\item We divide the needlet coefficients by their standard deviation due to CMB and noise to get the normalized needlet coefficients.
\item We loop on detection threshold starting with 50 sigma and gradually going down to 5 sigma.
\item For a given threshold, we loop on the pixels of the $N_\mathrm{side}=512$ map of normalized needlet coefficients. When a pixel with a value above the threshold is found, we identify a disc of radius equal to the distance of influence for the given needlet around this pixel. Possible amplitudes of the point source are estimated using as possible source positions the centers of all pixels in a $N_\mathrm{side}=2048$ map within this disc. Thus for all $N_\mathrm{side}=2048$ pixels within the disc we obtain the amplitude of the point source assuming that each of these pixels are the centers. The center of the pixel which gives the highest estimate of the amplitude is identified as the most likely position of the source.
\item Using the best fit source center, we subtract the best fit point source model from the $N_\mathrm{side}=512$ needlet map. This is done in order to avoid further detections of the same source as we continue looping through the pixels.
\item After the loop on pixels and detection thresholds, we are left with a list of positions and amplitudes.
\item Finally we need to identify the cases where residual diffuse foregrounds, close sources or extended sources give rise to a false detections or detections of sources where we are unable to estimate a reliable amplitude. We use a $\chi^2$ goodness of fit test to eliminate these detections from the list. For simplicity (and reduced CPU time), we ignore correlations between needlet coefficients in the $\chi^2$ and use a diagonal covariance matrix. In order to determine a limiting $\chi^2$ above which we do not accept the detection, we look at the sources starting with the highest $\chi^2$ values and continue until we reach values where we see that the point sources have reasonable shapes. We also eliminate some real point sources in this step, but the number turns out to be so small that we can justify this procedure. The reduced $\chi^2$ limits are listed in table \ref{tab:chi-lims}.
\item A few very bright sources give rise to two detected sources very close to each other due to imperfect subtraction of the source upon the first detection. Both of these sources tend to give acceptable $\chi^2$ and are not rejected by the $\chi^2$ test. In order to reduce these to one detection, we go trough all sources and look for those which are closer than an identification radius of $0.4^\circ$ (As will be shown below, this is the $5\sigma$ error on the distance between two sources). For sources which are closer, only the one with the highest estimated amplitude is kept.
\item We have now obtained a set of detected point sources with positions and amplitudes.
\item For sources which are detected only in internal templates and not directly in the channels, we cannot estimate the amplitude from the template. The template only allows for finding the difference in amplitude between channels. In this case, we use the position found in the template to search for the best fit position in the individual channels where we then estimate the amplitude. If the amplitude is non-zero at the $3\sigma$ level or more (note that the source was detected at the $5\sigma$ level in the template), we count the source as detected in the channel with a reliable amplitude and position.
\end{enumerate}

\begin{deluxetable}{lcc}
\tabletypesize{\normalsize}%\tiny}%\small}%normalsize}%\tiny}%scriptsize}%\footnotesize}
\tablecolumns{3}
%\tablewidth{0pt}
\tablecaption{Limits on $\chi^2$ valid both for simulations and WMAP 7 year data\label{tab:chi-lims}}
\tablehead{   % column headings
 \colhead{Needlet}  &
\colhead{$F_s$ [Jy]/ $T_S$ [mK]\tablenotemark{a}} &
   \colhead{LIMIT}
}
\startdata
 Mex B=1.9 j=9 (K)         &$F_S <1.91$ [Jy]  & 2.68\\
Std B=1.8 j=10 (Ka)        &$F_S <3.20$ [Jy]  & 2.60\\
Std B=1.6 j=13 (Q)         &$F_S <2.77$ [Jy]  & 2.43\\
 Std B=2 j=9   (V)         &$\forall \; F_S$  & 2.10\\
Mex B=1.8 j=11 (W)         &$\forall \; F_S$  & 1.76\\
Mex B=1.9 j=8  (K-Ka,K-V,K-W) & $T_S<2.54\cdot10^{-1}$ & 2.00\\
                           & $T_S<3.66\cdot10^{-1}$ & 2.41\\
Mex B=1.8 j=9  (Ka-V)       & $T_S<2.68\cdot10^{-1}$ & 1.89\\
                           & $T_S<5.36\cdot10^{-1}$ & 2.31\\
Mex B=2 j=8    (Q-V)        & $T_S<3.35\cdot10^{-1}$  & 1.76\\
                           & $T_S<4.18\cdot10^{-1}$ & 2.01\\
\enddata
\tablecomments{The $\chi^2$ acceptance limits are only valid for sources with a flux (respectively temperature) lower than those given in the table. For larger fluxes/temperatures, we need to do by-eye inspection to separate extended foregrounds from a strong point source}
\tablenotetext{a}{For the 5 WMAP-channels the limit until where the $\chi^2$-criterion is valid is given by a limiting flux in units of Jansky, while for templates by a limiting temperature. This is a consequence of the fact that in templates we measure a difference in temperature between two channels at different frequencies and hence there is no obvious way to transform such an amplitude to Jansky. }
\end{deluxetable}

\section{Results on simulations}
\label{sec:simul}

\subsection{Creating the simulations}

The aims of the simulations are:
\begin{itemize}
\item to test if the estimates of source amplitudes are unbiased
\item to find error bars on amplitude and position of the sources
\item to find the detection limits for the different channels and templates
\item to identify problems with the detection algorithm
\end{itemize}

We thus try to make simulations which are simple, fast and mainly fitted to fulfill these goals more than to make simulations with realistic point source amplitudes and numbers of sources. But in order to have a range of amplitudes which are not too unrealistic, we choose to use 464 sources detected in a first run in the WMAP 7 year K-band data. To test the lower limit of detection we added 71 sources slowly decreasing in amplitude from the smallest of the 464 K-band sources.

We simulate the positions in such a way that the minimum distance between the sources is always larger than $1^\circ$. As we show later, there are some problems with the detection and estimate of the amplitude of sources which are very close. In order to obtain reliable error estimates from simulations, we need to reduce this problem here by simulating source positions which are more than $1^\circ$ apart. The centers of the sources are taken to be centers of pixels at HEALPix resolution $N_{side}=2048$. The amplitudes of the sources on other channels than the K channel are obtained assuming a synchrotron spectral index  of $-2.7$. Again, this is not completely realistic but sufficient to satisfy the goals of the simulations. We now make a pure point source map with the above defined positions and amplitudes.

Finally we make 3000 different realisations of noise and CMB fluctuations and add the pure point source map to them, obtaining maps with CMB, noise and point sources, but no extended foregrounds. To generate CMB and noise realizations, we use the best fit WMAP7 power spectrum and the WMAP noise rms models. The choice of the number of simulations was motivated by the trade-off between available CPU time and the necessary accuracy on the error bars obtained from simulations.

\subsection{Analyzing the simulations}
\label{sec:analsim}

First we used 3000 simulated maps to estimate the error $\sigma^\mathrm{Pos}$ on the estimated position.   Only input sources which are detected in at least 1000 simulations are used to estimate error bars.  In order to identify a detected source with an input source when estimating the error on position, we needed to ensure that we used a search radius which was much larger than the $1\sigma$ error on position. An identification radius of 0.5 degrees was found to be sufficient.  After the error bars on position have now been found, we will in all further analysis of the simulations use a new identification radius of $0.25^\circ$ for the individual channels and the Ka-V and Q-V templates and $0.3^\circ$ for all other internal templates. This corresponds to a maximum of about $5\sigma^\mathrm{Pos}$ (taking the error for the weakest sources); very few sources are expected to be found outside a radius of $5\sigma$. For some channels the $5\sigma^\mathrm{Pos}$ distance is less than 0.25 degrees, we still use 0.25 degrees as identification radius in order to include at least 2 pixels on $N_\mathrm{side}=512$. When deciding whether a source detected in one channel may be the same source as one detected in another channel at a slightly different position, we will use a maximum distance of $\sqrt{2}\times5\sigma\approx0.4^\circ$ for identification talking into account the $5\sigma$ error on position for both sources.

We then used 3000 simulated maps to estimate the error $\sigma^A$ on estimated amplitude now using the new identification radius. We find that the error on the amplitude is independent of its value (but fluctuates around a constant value, due to noise) whereas the error on position grows with decreasing amplitude. To be conservative, we will use the error bars for the weakest sources. The mean error bars on amplitude, the mean value of the error bars on position as well as the error bars on the position for the weakest sources are all shown in table \ref{tab:channel} for individual channels and table \ref{tab:template} for the templates (in the latter table we do not list an error on the amplitude, since this is an amplitude difference between two channels which is never used).

\begin{table}[!hp]
\centering
\caption{Some results for the individual channels, based on 3000 simulations}
\begin{tabular}{|c|c|c|c|c|c|}
\hline
WMAP-channel $i$                           &  K      & Ka       & Q        & V         &W \\
\hline
$\langle\sigma^{A}_i\rangle [\mathrm{Jy}]$\tablenotemark{a} &0.165&0.158 &0.173 &0.225 &0.315 \\
\hline
$\sigma^{Pos}$ [arcmin]                    &3.03     & 2.88     & 1.74     & 1.40      &1.24\\
$\langle\sigma^{Pos}\rangle$ [arcmin]      &1.87     & 1.57     & 1.29     & 1.18      &0.95\\
\hline
Found total (535 input sources)            &424      & 353      & 216      & 87        & 24\\
After $\chi^2$ elimination                              &420      & 351      & 215      & 87        & 24\\
Identified                                 &419      & 350      & 213      & 85        & 22\\
Unique                                     &82       & 12       &  1       & 0         & 0 \\
Avg false positives\tablenotemark{b}       &1.7      &1.6       & 1.9      &1.8        & 1.7\\
Detected in int.temp                       &462      &448       &373       &191        &65\\
\hline
Amplitude detection limit [Jy]             &0.519&0.501 &0.491 &0.522 &0.991 \\ 
\hline
Avg 99\% completeness flux, channels [Jy]\tablenotemark{c} &1.07&1.59 &1.12 &1.40&1.79 \\
\hline
Avg 99\% completeness flux, int. temp. [Jy]\tablenotemark{c} &1.01&0.883 &0.932 &1.09&1.46 \\
\hline
Avg 99\% completeness flux, combined [Jy]\tablenotemark{c} &0.775&0.747 &0.830 &1.08&1.46 \\
\hline
\end{tabular}
\label{tab:channel}
\tablenotetext{a}{NB: this error on the amplitude does not take into account the error of the effective area (used to convert Kelvin to Jansky), since this error is dependent on the value of the amplitude. }
\tablenotetext{b}{Represents the average number of detections not identified with an input source, discrepancies from ``After $\chi^2$ elimination'' minus ``Identified'' come from rounding.}
\tablenotetext{c}{Represents the average input flux limit from where 99\% of the simulated input sources are detected. ``Channels'' standing for the sources detected directly in the 5 channels; ``int. temp.'' for those detected in internal templates and being non-zero in channels at the $3\sigma$ level; ``combined'' those detected in either the channels or the templates.}
\end{table}

\begin{table}[!hp]
\centering
\caption{Some results for the internal templates, based on 3000 simulations}
\begin{tabular}{|c|c|c|c|c|c|}
\hline
template $i$                                       &  K-Ka     & K-V       & K-W       & Ka-V       &Q-V \\
\hline
$\sigma^{Pos}$ [arcmin]                            &  3.46     & 3.31      &3.34       & 2.94       & 2.65\\
$\langle\sigma^{Pos}\rangle$ [arcmin]              &  2.00     & 1.36      & 1.36      & 1.67       & 2.06\\

\hline
Found total (535 input sources)                    & 494       & 550       &  549      & 481        &288\\
After $\chi^2$ elimination                                      & 444       & 458       &  460      & 456        &285\\
Found Identified                                   & 442       & 457       &  459      & 455        &282\\
Unique                                             & 1         & 1         &  0        & 6          & 5 \\
Avg false positives\tablenotemark{a}           &1.8  &1.0   & 1.1  &1.8    & 2.2\\

\hline
amplitude difference, det. limit [$10^{-2}mK$]\tablenotemark{b}     & 3.96      &4.67       &4.98       & 5.17       &5.76 \\
\hline
\end{tabular}

\label{tab:template}
\tablenotetext{a}{Represents the average number of detections not identified with an input source, discrepancies from ``After $\chi^2$ elimination'' minus ``Identified'' come from rounding.}
\tablenotetext{b}{As previously, the minimum amplitude difference is reported in temperature units rather than Jansky. This is a consequence of the fact that in templates we measure a temperature difference between two channels at different frequencies and hence there is no obvious way to transform such an amplitude to Jansky. }
\end{table}

We compare the mean value of the estimated amplitudes in the simulations with the corresponding input amplitude. We find that the estimated amplitudes are unbiased with exception of the very weakest sources which are influenced by the Eddington bias (\cite{edd}). In section \ref{sec:wmap} we will show how we correct for the Eddington bias in the WMAP data.

In table \ref{tab:channel} (for individual channels) and table \ref{tab:template} (for templates) we show the mean number of sources found in the 3000 simulated maps (entry ``Found total''). We list both the number of $5\sigma$ detections as well as the mean number of sources remaining after the applying the $\chi^2$ acceptance criterion (entry ``After $\chi^2$ elimination''). The mean number of these accepted sources which are identified with input sources is also shown (entry ``Identified''). The detections which are not identified are found to be either random fluctuations or sources which input position is further than $5\sigma^\mathrm{Pos}$ away from the detected position, they are shown in entry ``Avg false positives''. We also show the mean number of unique detections (entry ``Unique''), for the individual channels this refers to sources which are detected only in one channel, for the templates this refers to sources detected only in one template and no others. A given source is a unique detection in the channels (respectively templates), if no other channel (respectively template) has detected a source within $0.4^\circ$ from the position of this source, where $0.4^\circ$ comes from the error on positions as explained above.

It may seem surprising that while the Ka-V and Q-V templates overall detect less sources than the other templates, they detect on average more unique sources, not detected in other templates. The reason for this is that the needlets used for these templates have smaller spatial extension and can hence better resolve point sources which are very close. This is also the reason why there are much less sources which are eliminated by the $\chi^2$ criterion in these two templates; the number of partially overlapping sources is much smaller.

In the same tables, we also show the mean (input) amplitude of the weakest detected sources in the simulations. This is the amplitude limit below which very few sources will be detected. For the internal templates, we show the weakest differences in amplitude rather than the amplitude itself.
In the 5 frequency channels, an average of 435 sources are found in total among all the channels, for the templates, the corresponding number is 491. Thus, on average 56 more sources are found in the templates; this indicates that the lower background level (pure noise versus noise+CMB) of the templates increases the number of detections, in spite of the fact that the amplitudes in the templates are differences and not absolute values. The simulations show that this is true for synchrotron sources, in real data we will see that this effect is even stronger and that using the internal templates gives the possibility to discover many more sources than when using only the individual channels.

We also follow up the point sources which are detected in the templates by amplitude measurements at the same position in the individual channels. We first use the two templates with the smallest error in position (Ka-V and Q-V). For all sources detected in these templates, we look for the maximum amplitude in a radius of 0.25 degrees in the 5 individual channels. For the sources detected only in the remaining 3 templates we use a radius of 0.3 degrees. The radii of 0.25 and 0.3 degrees correspond again to roughly 5 times the error on the position as explained above. In order to make sure that the given source is actually seen in a given channel, we only accept the detection and use the amplitude if it is non-zero at least at the $3\sigma$ level. In table \ref{tab:channel} we show the number of sources detected by this method in each of the channels.

Finally, we also show in table \ref{tab:channel} the average limiting flux from where the detection is 99\% complete, meaning from where we detect in average 99\% of the source with a greater flux than the reported limit. We report this average completeness limit for the detections directly done in the five WMAP channels, for $5\sigma$ detections in the templates which are non-zero at least at the $3\sigma$ level in the channels and for the combined unique detections of the two preceeding approaches. This shows well how the two approaches are complementing each other.

\section{Results on WMAP 7 year data}
\label{sec:wmap}

\begin{table}[t!h]
\centering
\vspace{0.2 cm}
\caption{Detection, acceptance and identification on WMAP 7-year data}
\begin{tabular}{|l|c|c|c|c|c|}
\hline
 {\bf TEMPLATE detected at $\geq5\sigma$}                          & {\bf K-Ka}             & {\bf K-V}            & {\bf K-W}            &{\bf Ka-V}  &{\bf Q-V}\\
\hline
Found total                                                  &3573                    & 7190                 &  6609                & 1017       & 341  \\
\hline
After $\chi^2$ elimination                                   &967                     & 1186                 &  1219                & 633        & 303  \\
\hline
Unique \tablenotemark{a}                                     &233                     & 247                  &  262                 & 138        & 24   \\
\hline
Identified in WMAP catal.                                    &  335                   & 287                  & 307                  & 393        & 287 \\
\hline
Identified in NEWPS\_5yr\_5s                                 &295                     &260                   & 278                  & 330        & 259  \\
\hline
Identified in ERSC                                           &    340                 &316                   &327                   &374         & 270      \\
\hline
New sources\tablenotemark{b}                                  &561                     &810                   &829                   &185         & 8 \\
\hline
Unique new sources\tablenotemark{c}                           &217                     &242                   &253                   &87          & 3      \\
\hline
 {\bf CHANNEL detected at $\geq5\sigma$}                     & {\bf K}                & {\bf Ka}             & {\bf Q}              &{\bf V}     &{\bf W}\\
\hline
Found total                                                  & 587                    &  511                 & 385                  & 167        &  57    \\
\hline
After $\chi^2$ elimination                                   & 471                    &  295                 & 267                  & 161        &  55    \\
\hline
Unique \tablenotemark{a}                                     & 174                    &  19                  & 17                   & 3          &  2   \\
\hline
Identified in WMAP catal.                                    & 404                    &  282                 & 263                  & 159        &  54    \\
\hline
Identified in NEWPS\_5yr\_5s                                 & 373                    &  274                 & 245                  & 156        &  52    \\
\hline
Identified in ERSC                                           & 371                    &  266                 & 252                  & 159        &  55    \\
\hline
New sources\tablenotemark{b}                                  & 40                     &  7                   &  3                   &  1         &  0     \\
\hline
Unique new sources\tablenotemark{c}                           & 40                     &  3                   &  2                   &  1         &  0     \\
\hline
{\bf CHANNEL $\geq3\sigma$, detect. in templ. $\geq5\sigma$} & {\bf K}                & {\bf Ka}             & {\bf Q}              &{\bf V}     &{\bf W}\\
\hline
Found total $\geq3\sigma$             &916                     & 642                  &576                   &356         &172  \\
\hline
Unique\tablenotemark{a}               &263                     &55                    &55                    &12          &8    \\
\hline
Identified in WMAP catal.             &455                     &423                   &407                   &292         &140  \\
\hline
Identified in NEWPS\_5yr\_5s          &400                     &366                   &352                   &270         &135   \\
\hline
Identified in ERSC                    &468                     &426                   &402                   &295         &144       \\
\hline
New sources\tablenotemark{b}          &360                     &150                   &115                   &37          &23 \\
\hline
Unique new sources\tablenotemark{c}   &221                     &50                    &48                    &12          &8   \\
\hline
\end{tabular}
\tablenotetext{a}{``Unique'' stands for all the sources uniquely detected in the given channel (respectively template), meaning no other detection within $0.4^\circ$ as for the simulations.}
\tablenotetext{b}{``New sources'' stands for all the sources found not in any of the 3 sets of catalogues.}
\tablenotetext{c}{``Unique new sources'' stands for all the sources found only in given template or channel and not in any of the 3 sets of catalogues.}

\label{tab:numbers}
\end{table}
 
In table \ref{tab:numbers} we show the results of the above procedure on the WMAP data. Note that we only search for sources outside the Kq85 galactic mask (to be exact, we use the WMAP point source catalogue mask which is similar but not equal to the Kq85 galactic mask (see \cite{WMAPsrc})). The table is divided in three parts, first we show the sources detection in the internal templates only, then in the individual channels only and finally the sources detected at $5\sigma$ in the templates and then found at more than $3\sigma$ amplitude in the channels. In each case we show the number of detections before (entry ``Found total'') and after elimination with the above $\chi^2$ criterion (entry ``After $\chi^2$ elimination''). For each channel and template we show how many sources are uniquely detected in this particular channel or template (entry ``unique'').

In the following, all references to detected sources refer only to those which passed the $\chi^2$ criterion. The table clearly shows the power of using the internal templates. The number of detected sources is substantially increased when including the templates. In total 522 sources were found using the channels only, whereas 2052 were found using the templates only. Overall, 1116 sources are detected in at least one frequency, either as $5\sigma$ directly on the map or as $5\sigma$ in the template and $\geq3\sigma$ at the corresponding position in the map.
While the number of detected sources in the templates is much larger than in the maps directly, there are still 50 sources which we detect in the channels but not in the internal templates. These are sources which are so close to other sources that they are not resolved by the relatively large extension of the needlets used in the internal templates; they can only be resolved by the sharper needlets used in the frequency channels.

 A bit more detail on how these numbers are obtained:
   \begin{description}
   \item[522 detections in the channels:] We start by taking all the detected sources in the K-channel (where we detect most), then we add those detected in the Ka-channel which are not also detected in the K-channel (meaning further than $0.4^\circ$ from the positions of the detections in K). Then we iterate till the W-channel. Note, this procedure just counts how many different sources we detect, independently of whether they are unique or detected in more than one channel.
   \item[2052 detections in templates:] Same approach as for the 522 detections in the channels, starting from the sources detected in Ka-V, then adding those not already present iteratively from Q-V, K-W, K-V and K-Ka.
     \item[1116 detections in at least one frequency :] As for the $5\sigma$ detections in the channels we combine the different sources we detect at $5\sigma$ in the templates and at $3\sigma$ in the channels. We keep all 522 sources detected directly in the channels and add those from the templates which are further than $0.4^\circ$ from the ones directly detected in the channels.
   \end{description}

Note that with a $5\sigma$ detection criterion, we would expect about 18 false detection in total considering that we have 5 channels and 5 templates. In the simulations we had an average of 9 false detections in the channels and 8 in the templates. Furthermore, for the sources which are detected at $5\sigma$ in the templates only, a total of 2052, one should expect about 6 false $3\sigma$ detections for each channel. We therefore expect about 30 false detections among the sources which are only detected at $5\sigma$ in the templates and  $3\sigma$ at the same position in the channels. We exclude some of these by noting that some detections in the V and W band are positive $3\sigma$ detections whereas at the same positions in the internal templates K-V and K-W there are positive $5\sigma$ detections, meaning that the source should be much stronger in K than in V or W. If at the same position, there is no $2\sigma$ detection in neither of the two other bands among Q, V and W, we assume that the detection is due to a fluctuation and is excluded. We find in total 30 such cases.

\subsection{Comparison with catalogues at the same frequencies}

In order to test our method we will first compare the sources to those found by the WMAP team using the WMAP 7 year PS (\cite{WMAPsrc}) catalogues, sources found in the NEWPS\_5yr\_5s catalogue (\cite{mass}) based on WMAP 5 year data and the Early Release Compact Source Catalogue (ERCSC) based on Planck observations (\cite{Planck}) at frequencies 30, 44, 70 and 100 GHz. The WMAP catalogues contain 542 independent sources, the  NEWPS\_5yr\_5s catalogues contains 533 and the ERCSC catalogues contain 705,452,599 and 1381 sources at respectively 30, 44, 70 and 100 GHz (adding up to 1585 distinct sources if we identify them amongst themselves when they are within the identification radius of $0.4^\circ$). In our work we have concentrated the search outside the Kq85 galactic mask, the corresponding number of sources found in these three catalogues are thus 536 (WMAP), 430 (NEWPS) and 678 (ERCSC).

Of the 536 sources in the WMAP catalogues, we find 487 (combining individual channels and templates) with our procedure. Of the 49 missing WMAP sources, 24 are detected but rejected by our conservative $\chi^2$ criteria, another 12 are detected but not identified as they are offset by more than our $5\sigma^\mathrm{Pos}$ identification radius. Only 13 of the WMAP sources are not detected at all. These are all very weak and close to other stronger sources and therefore not resolved by the needlet coefficients. For the NEWPS catalogue, we find 415 of the 430 sources, the remaining 15 sources are detected but excluded by our $\chi^2$ criterion. Among the 678 ERCSC sources, we detect in total 517 by our procedure. For the 1116 sources which are found either directly at $5\sigma$ in the channels or at $5\sigma$ in the templates and $3\sigma$ in at least one channel, 506 are new sources which are not listed in the WMAP, NEWPS or ERCSC catalogues. In table \ref{tab:numbers} we show channel by channel the number of sources identified with sources in these three catalogues (entries ``Identified in WMAP catal.'', ``Identified in NEWPS\_5yr\_5s'' and ``Identified in ERSC'').

\subsection{Comparison with catalogues at other frequencies}

For all these 1116 sources we run the identification procedure with catalogues at other frequencies. In table \ref{tab:new-PS} we show the relevant data. The first column corresponding to the source number, the second and third to right ascension (J2000) and declination in degrees, given by the weighted average of the positions in all frequencies where the source is found. Columns 4 to 8 correspond to the flux densities estimated at the given frequency. For converting the  source temperature and errors from Kelvin to Jansky we use the effective beam area and relative error as presented in table 4 of \cite{effarea}. For the very weakest sources, we are unable to correct the flux for Eddington bias due to the low signal-to-noise level (see section \ref{sec:spanbiascorr} below). For these sources we use the internal template to estimate the flux as explained in section \ref{sec:spanbiascorr}. These fluxes are in italic in the table. In the 9th column there are 5 flags, one for each channel, indicating whether we detected the source at $5\sigma$ in the individual channel (``C''), or in the template with a $3\sigma$ detection in the channel (``F'') or not at all (``.''). In the 10th column there are flags indicating if the found source is identified with one in a given catalogue in at least one frequency: ``W'' for WMAP-catalogues, ``N'' for the NEWPS\_5yr\_5s-catalogue and P for the ERCSC-catalogues (at 30,44,70,100 GHz only). This table is also available at \verb=http:folk.uio.no/frodekh/PS_catalogue/Scodeller_PS_catalogue.txt=.

The last 2 columns contain the identification with the GB6 (\cite{GB6}), PMN (\cite{pmn1,pmn2,pmn3,pmn4}), 1Jy \citep{kuehr} when possible or with the lower frequency catalogues NVSS (\cite{nvss}) and SUMSS \citep{sumss} catalogues counterparts, the offset from the position in the given catalogues, a flag ``M'' if there are multiple identifications possible (where we took the brightest one) and a flag ``A'' if the listed counterpart has a flux density below 100 mJy.

Of the 1116 detected sources, 1021 have a 5 GHz counterpart (either GB6 or PMN), counting also the sources identified with ones from \citep{kuehr} there are 1029 having a rather high frequency counterpart. 69 sources have only a counterpart in the NVSS (61) or SUMSS (8) catalogues and 16 sources have no known\footnote{where ``no known'' means that even when there are counterparts in NVSS or SUMSS, but which are below 20 mJy we do not consider it.} counterpart.
Some of the identifications with weak sources in the catalogues (i.e the ones with flag ``A'') may be misidentifications. There are in total 104 sources with this flag. The mean distance between the position of our sources and the counterparts in catalogues is 12.1'. Excluding the sources with a weak counterpart (with flag ``A'') the mean distance to the counterpart for the remaining 1012 sources becomes 3.7' , substantially lower than the upper limit for identification (15').

Among the 16 found sources which have no known counterpart  14 are unique (meaning detected only in one channel) and all 16 new (meaning not in the WMAP, NEWPS\_5yr\_5s or ERCSC-catalogues), 6 of them are detected in K, 4 in Ka and 1 in Q, 5 in V and 2 in W. With the numbers in mind of false detections due to Gaussian fluctuations at $5\sigma$-level ($\approx 1.8$ for $N_{side}=512$), these 16 are most likely spurious detections due to Gaussian fluctuations.

Of the 506 new detections (meaning not in the WMAP, NEWPS\_5yr\_5s or ERCSC-catalogues) (235,195,8,52) have a counterpart in respectively GB6, PMN, SUMSS and NVSS catalogues and 46 of these have an identification with a weak counterpart (flag ``A'').  The mean distance to the counterpart for the new detections only is also 12.1' and if excluding the 46 new sources with flag ``A'', it becomes 8.1'. When considering only the sources which do not have flag ``A'' we see that the mean distance to the counterpart is larger when using only new sources than when using all detected sources. This is to be expected since the new detections tend to be rather weak (most of them are less than $5\sigma$ in the channels) and hence resolve the position worse.

\subsection{Bias correction}
\label{sec:spanbiascorr}
Flux-estimation is subject to the Eddington-bias \citep{edd}, which has the effect that with a given detection threshold the amplitudes of the weakest sources are over-estimated. Using a power law model ($\propto S^{-1\cdot(1+q)}$ for $S>S_m$ with S being source flux and $S_m$ a minimum flux from where the power law is valid) for the differential number count of galaxies,\cite{bias} present a Bayesian approach to correct for the bias. Their procedure allows to estimate both the slope of the power law as well as the bias.
The method is in two steps:
\begin{itemize}
\item estimate the slope via equation:
  \begin{equation}
    \label{eq:bias_q}
    \frac{1}{q}=\frac{1}{N}\cdot\(\sum_{i=1}^N\(\ln\(\frac{S_i^o}{S_m^o}\)+\ln\(\frac{1+\sqrt{1-4(1+q)/r_i^2}}{1+\sqrt{1-4(1+q)/r_m^2}}\)\)\),
  \end{equation}
where $S_i^o$ respectively $S_m^o$ are the observed (and hence biased) fluxes respectively minimum flux (which depends on the threshold) and $r_i$ is the observed signal to noise ratio of the source.

\noindent
This equation can be solved numerically if $r_m^2 \geq 4(1+q)$.

\item Then obtain the unbiased fluxes $S_i$ via:
  \begin{equation}
    \label{eq:bias_co}
    S_i=\frac{S_i^o}{2}\(1+\sqrt{1-\frac{4(1+q)}{r_i^2}}\).
  \end{equation}

\end{itemize}

For sources detected directly at $5\sigma$ in the WMAP-channels we find slopes 1+q respectively (2.04,2.07,2.06,2.14,2.12), while for the sources detected in the templates and then found at$\geq3\sigma$ in the channels (2.04,2.08,2.06,2.15,2.22). The two approaches give values of q in good agreement with each other and with other estimates see for instance, on WMAP-data \citep{Chen09} or \citep{lopez} and other rather high frequency surveys such as ATCA 18 GHz survey (\cite{ricci}), 9C survey at 15GHz (\cite{waldr}) or 33GHz VSA survey (\cite{Cleary}).

For sources with very low signal-to-noise ratio (in total for 71 different flux-estimations corresponding to 71 different sources ), the likelihood does not have a peak  ($r_m^2 < 4(1+q)$) and equation \ref{eq:bias_q} does not have a solution. In this case we are unable to estimate the bias and therefore also the flux. In order to solve this problem, we use the internal templates where the source has a much higher signal-to-noise ratio. To find the flux in channel $c$ where it is detected, then, if the source is not detected in channel $c'$ (where the distance in frequency between $c$ and $c'$ is as large as possible) we assume that the flux is zero in $c'$. With this assumption, we can use the amplitude estimated in the internal template and a corresponding bias correction to obtain the amplitude. To test this approach, we used a set of weak sources where we are able to estimate the bias correction with equation \ref{eq:bias_q}. For these sources, we estimated the amplitude both directly in the channel and using the internal template approach and found full agreement (within error bars) between the two methods. Nevertheless, in table \ref{tab:new-PS} we have written the fluxes estimated only from the internal templates in italic.

\subsection{Some examples}

\begin{figure}
\begin{center}
\leavevmode
\psfig {file=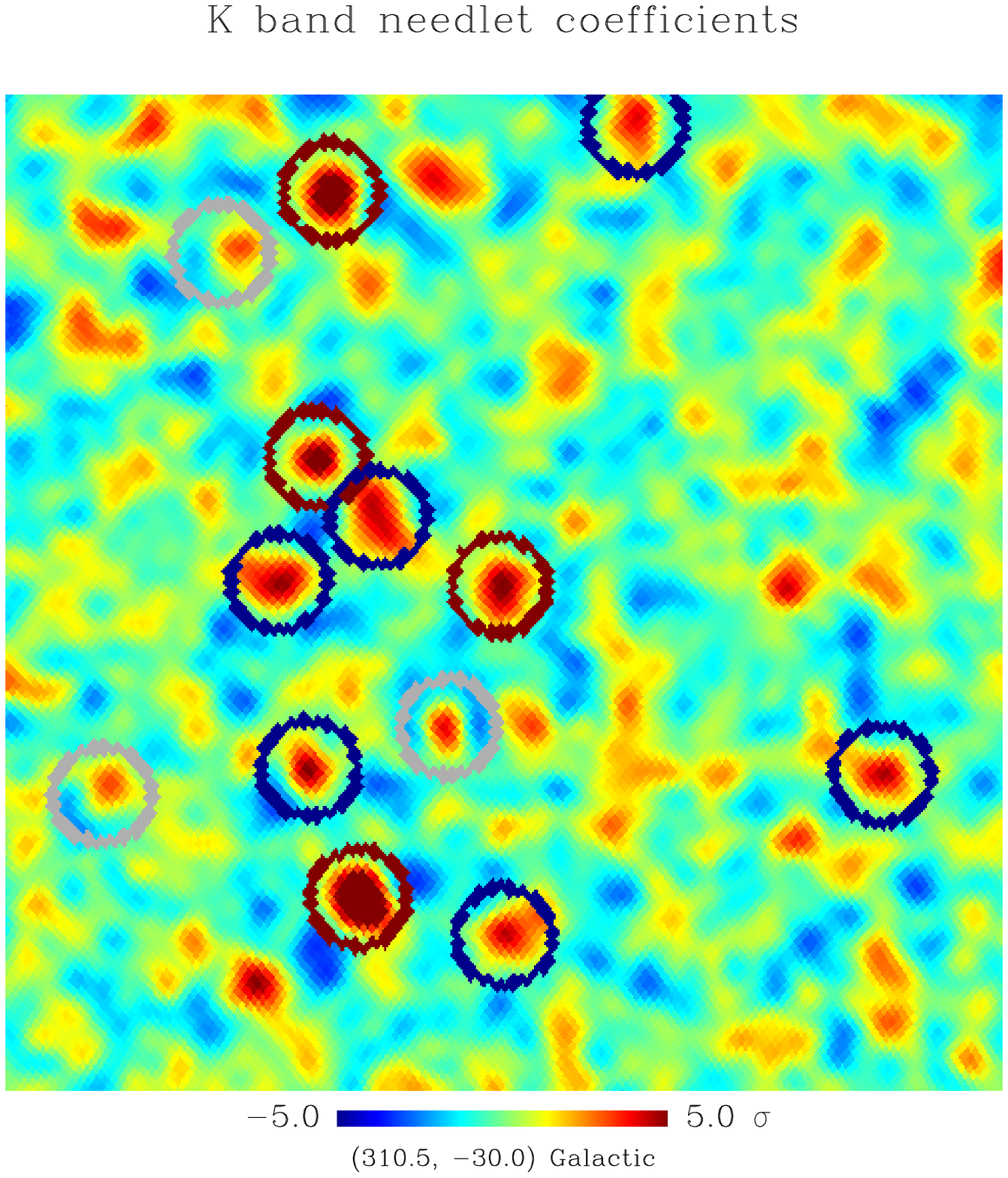,height=8cm,width=8cm}
\psfig {file=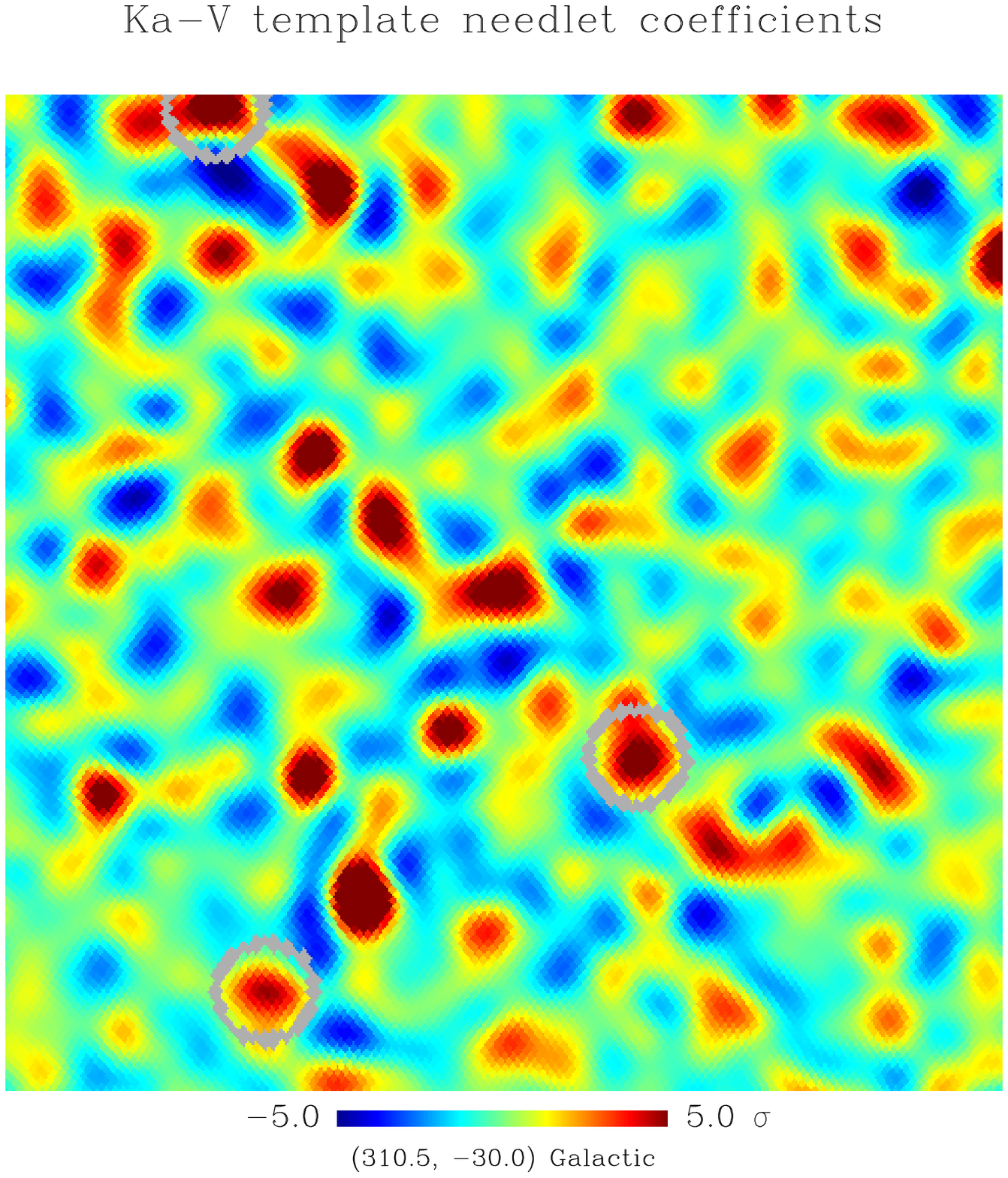,height=8cm,width=8cm}
\psfig {file=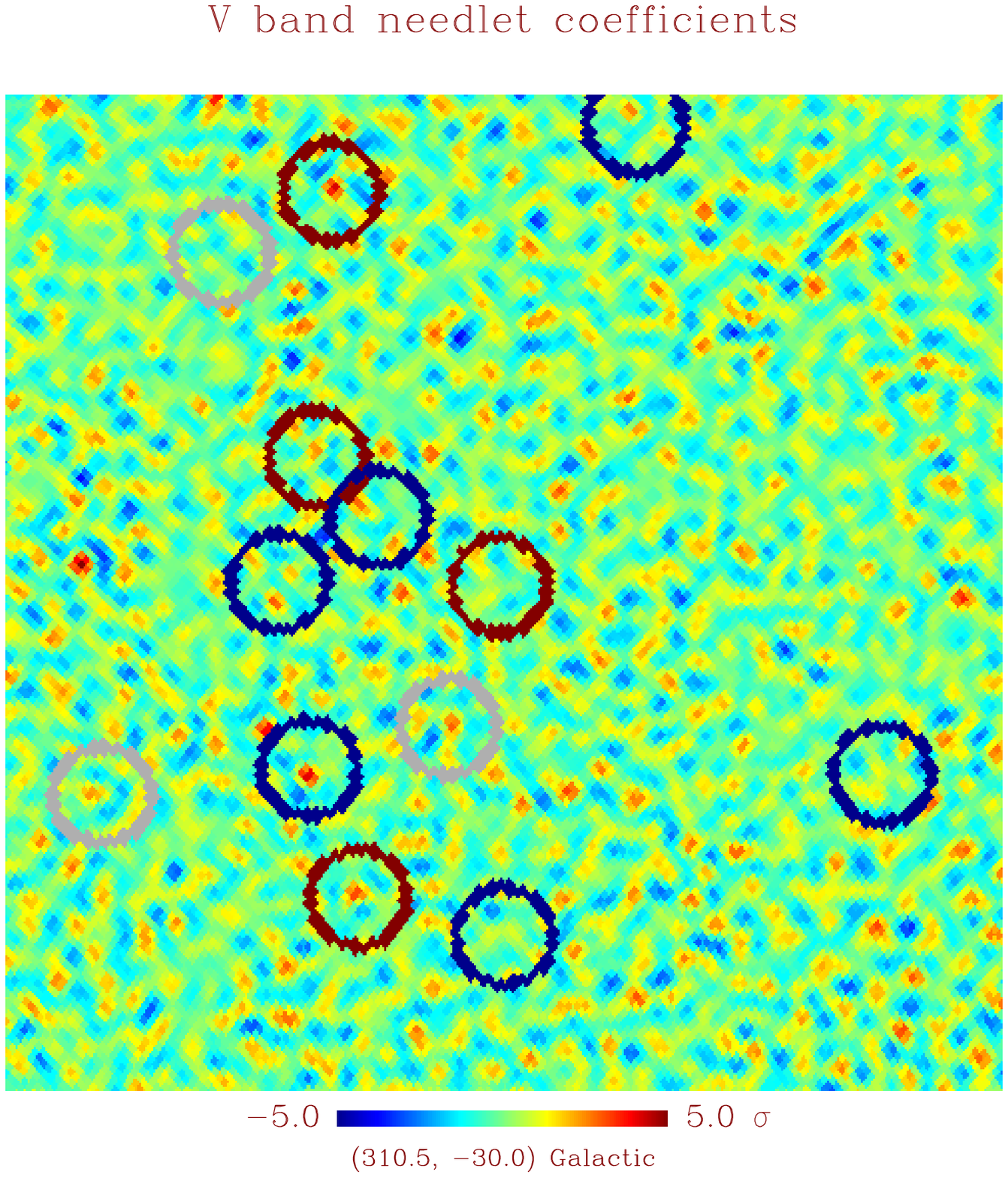,height=8cm,width=8cm}
\psfig {file=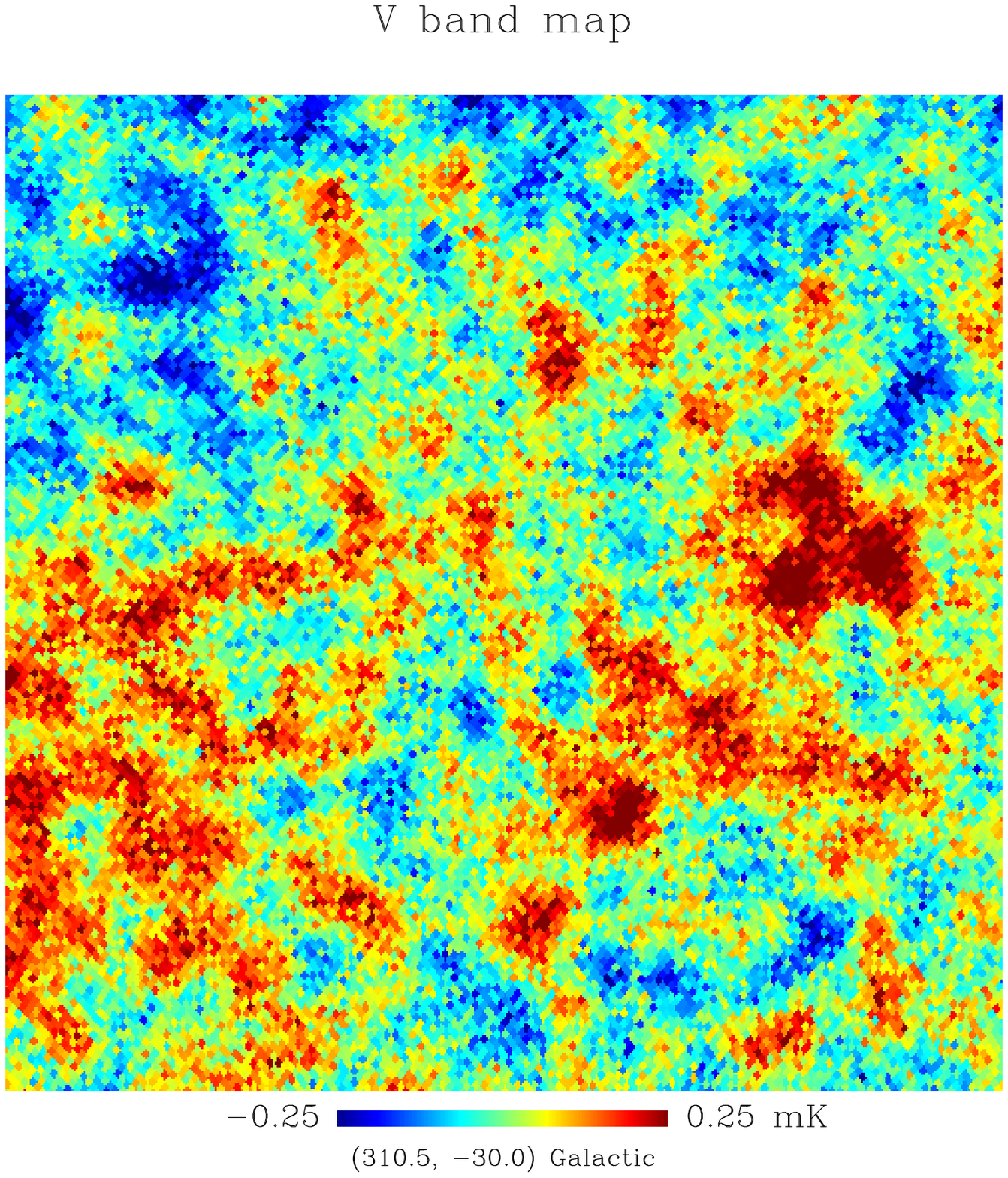,height=8cm,width=8cm}
\caption{All four plots show the same region on the sky. We show the needlet coefficients of the K and V band maps as well as the Ka-V template. We also show the V band map in real space. In the band maps, red circle indicates source at $\geq5\sigma$ in the K band, blue circle indicate $\geq5\sigma$ in a template and $\geq3\sigma$ in the K channel and grey circle means $\geq5\sigma$ in a template and $<3\sigma$ in the K channel. In the internal template, the three grey circles indicate sources which are $\geq5\sigma$ in the template but rejected (considered extended source/foreground) by the $\chi^2$ test.}
\label{fig:src}
\end{center}
\end{figure}

In figure \ref{fig:src} we show 4 projections, all of the same part of the sky. The needlet coefficients for the K and V band maps as well as the internal template Ka-V are shown. For reference we also show the V band temperature map. The sources which are detected (i.e. above $5\sigma$) in the K band are marked with a red circle, the sources which are detected ($\geq5\sigma$) in a template only and found $\geq3\sigma$ in the K band are marked with a blue circle. Finally the sources which we detect ($\geq5\sigma$) in the template but which are below $3\sigma$ in the K band are marked with grey circles in the K and V band needlet maps. Comparing the K band coefficients with the coefficients of the template we see clearly how many more sources are seen in the template. Note also that some of the sources detected in the template seem off center in the K band circles. The reason for this is that the circles are made based on the centers detected in the template which due to fluctuations are often displaced with regard to the frequency maps. Looking at the V band coefficients, we see that several of the detected sources are still present but at a much smaller amplitude.

\begin{figure}                                          
\begin{center}                                          
\leavevmode                                             
\psfig {file=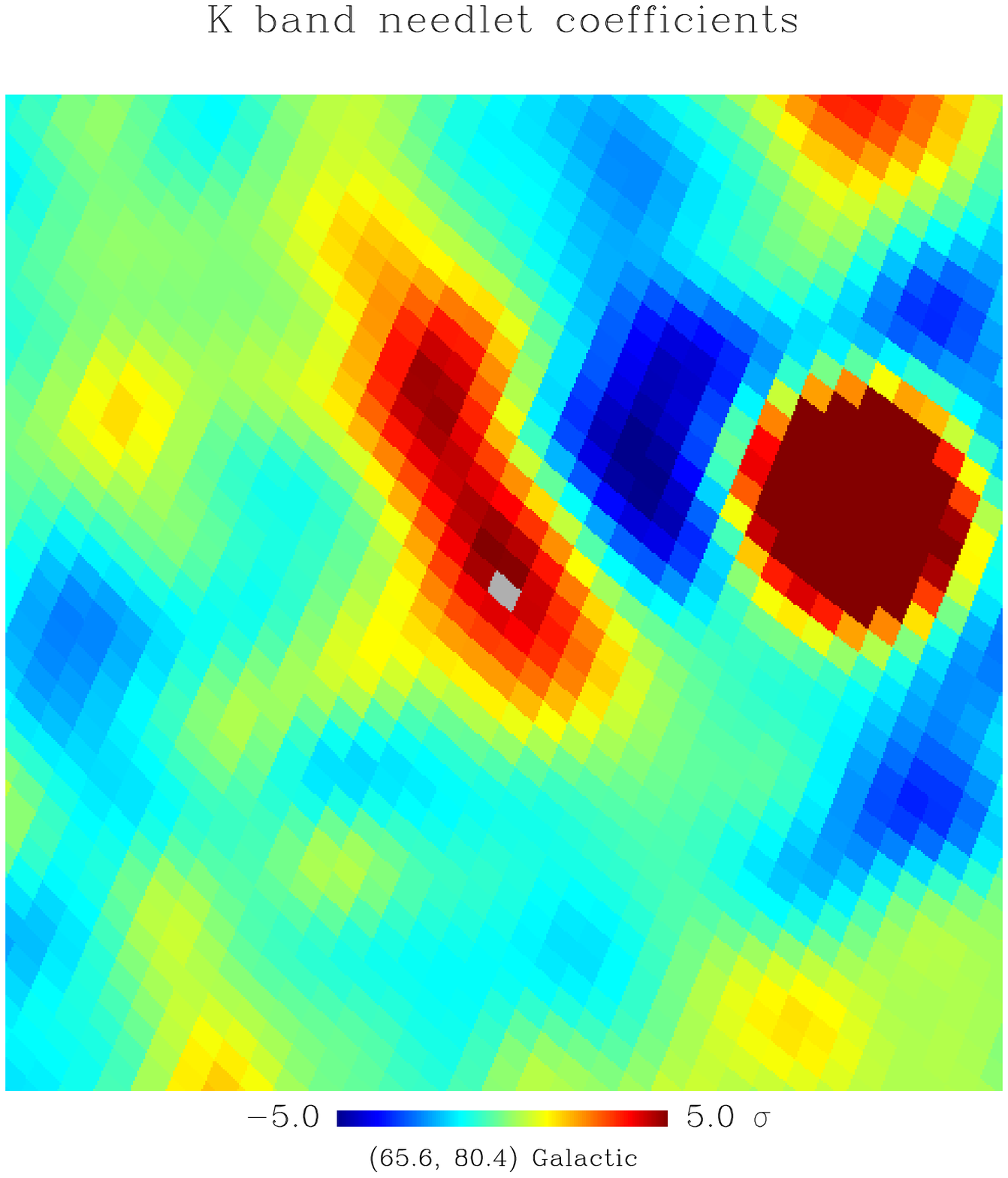,height=8cm,width=8cm}
\psfig {file=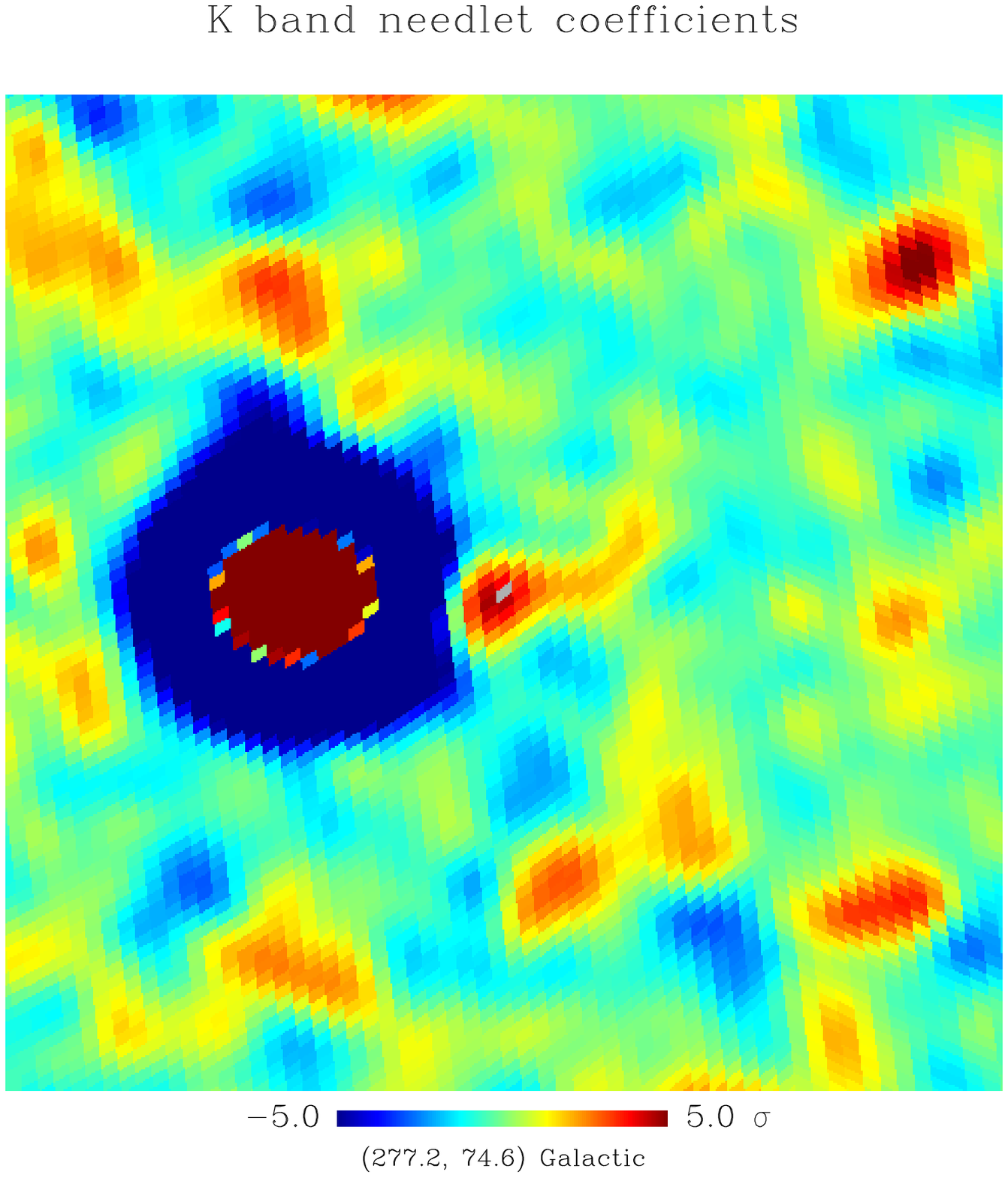,height=8cm,width=8cm}
\caption{The two figures show the needlet coefficients of sources in the K band in two different parts of the sky. Grey dot indicates the source center as detected by the WMAP team. In our procedure these two sources are detected but rejected by the $\chi^2$ test.}
\label{fig:src2}
\end{center}
\end{figure}

In the figure showing the needlets coefficients for the internal template there are three grey circles. These indicate sources which are detected $\geq5\sigma$ in the template but then rejected by the $\chi^2$ test. We can clearly see how some of these appear more elongated than the accepted sources and according to the $\chi^2$ test these do not resemble the beam shape; this may indicate that the sources are extended, but also simply that the mean beam shape is not a good approximation in this part of the sky. In figure \ref{fig:src2} we show two more examples of sources which are detected but rejected by the $\chi^2$ test. Both of these are detected by the WMAP team; the grey dot indicates the center as detected by WMAP. In the first case there seem to be two similar but very close sources, producing the elongated shape which the $\chi^2$ interprets as not being a point source. In the second case there is weak source close to a very strong one; the needlet amplification of the strong source distorts the vicinity of the weak source, making it fail the $\chi^2$ test.

\begin{figure}                                          
\begin{center}                                          
\leavevmode                                             
\psfig {file=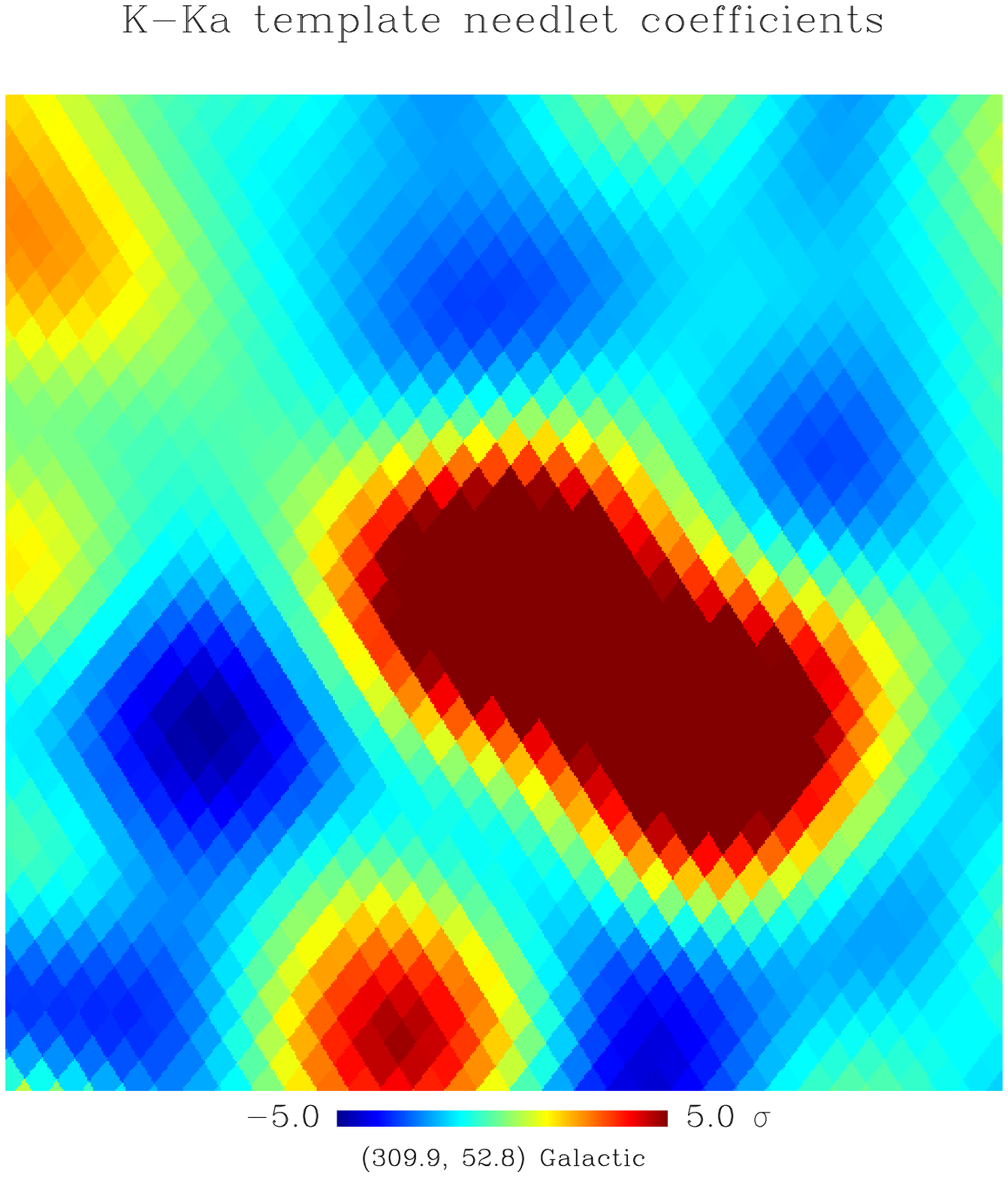,height=8cm,width=8cm}
\psfig {file=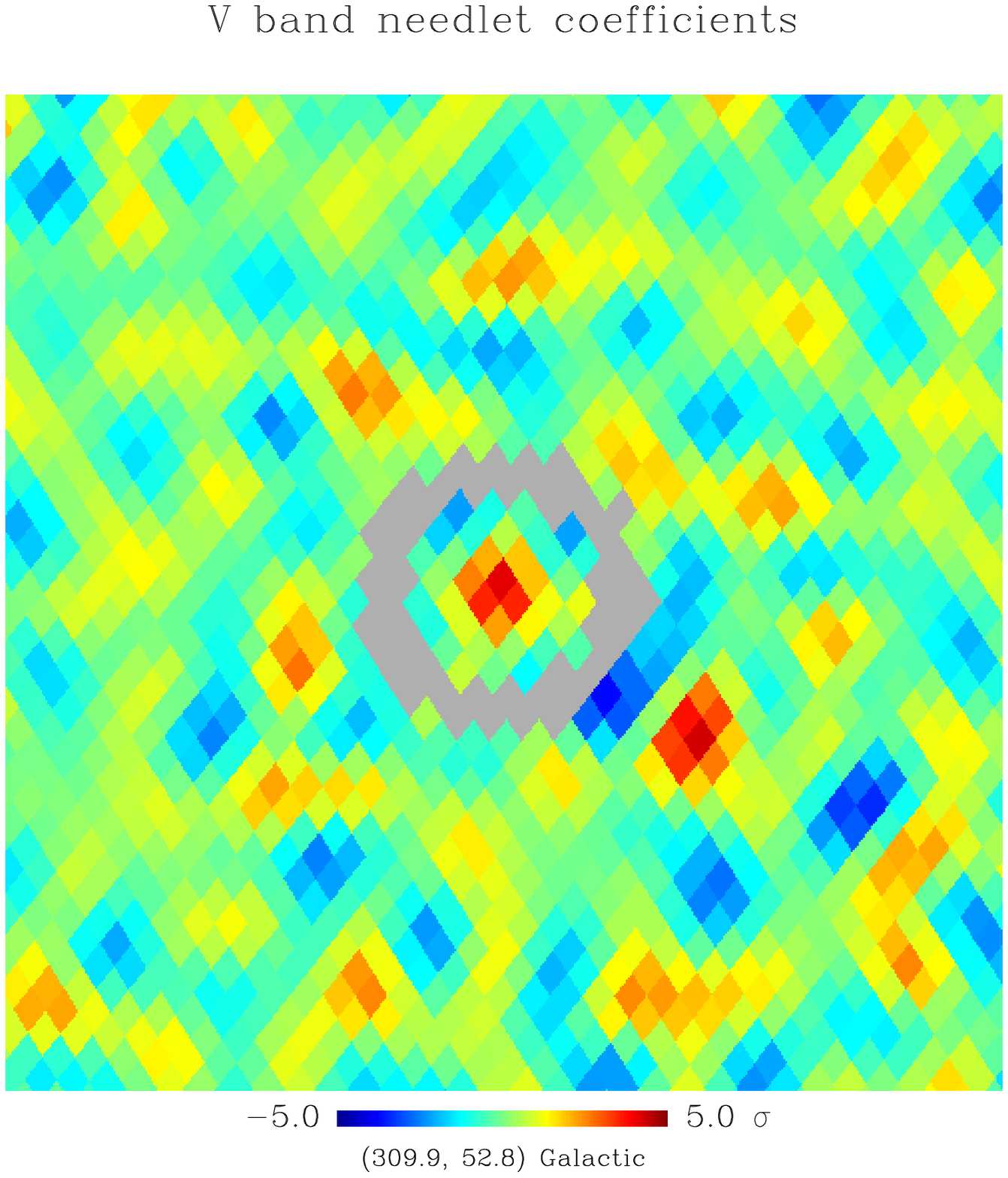,height=8cm,width=8cm}
\caption{The two projections show the same part of the sky. In the center we see one of our new sources (not found in other catalogues taken at the same frequencies, but identified with a source in catalogues at different frequencies) detected in the internal templates and found at $\geq3\sigma$ in several frequency channels.}
\label{fig:src3}
\end{center}
\end{figure}

Finally in figure \ref{fig:src3} we show an example of one of our new sources which are not listed in the WMAP, NEWPS or ERCSC catalogues, but identified with a 5GHz counter part (in PMN). The source is clearly seen above $5\sigma$ in the K-Ka template, but in the figure we see that it is also above $3\sigma$ in the V band needlet coefficients.

\begin{figure}                                          
\begin{center}                                          
\includegraphics[trim = 7mm 1mm 4mm 0mm, clip,scale=0.85]{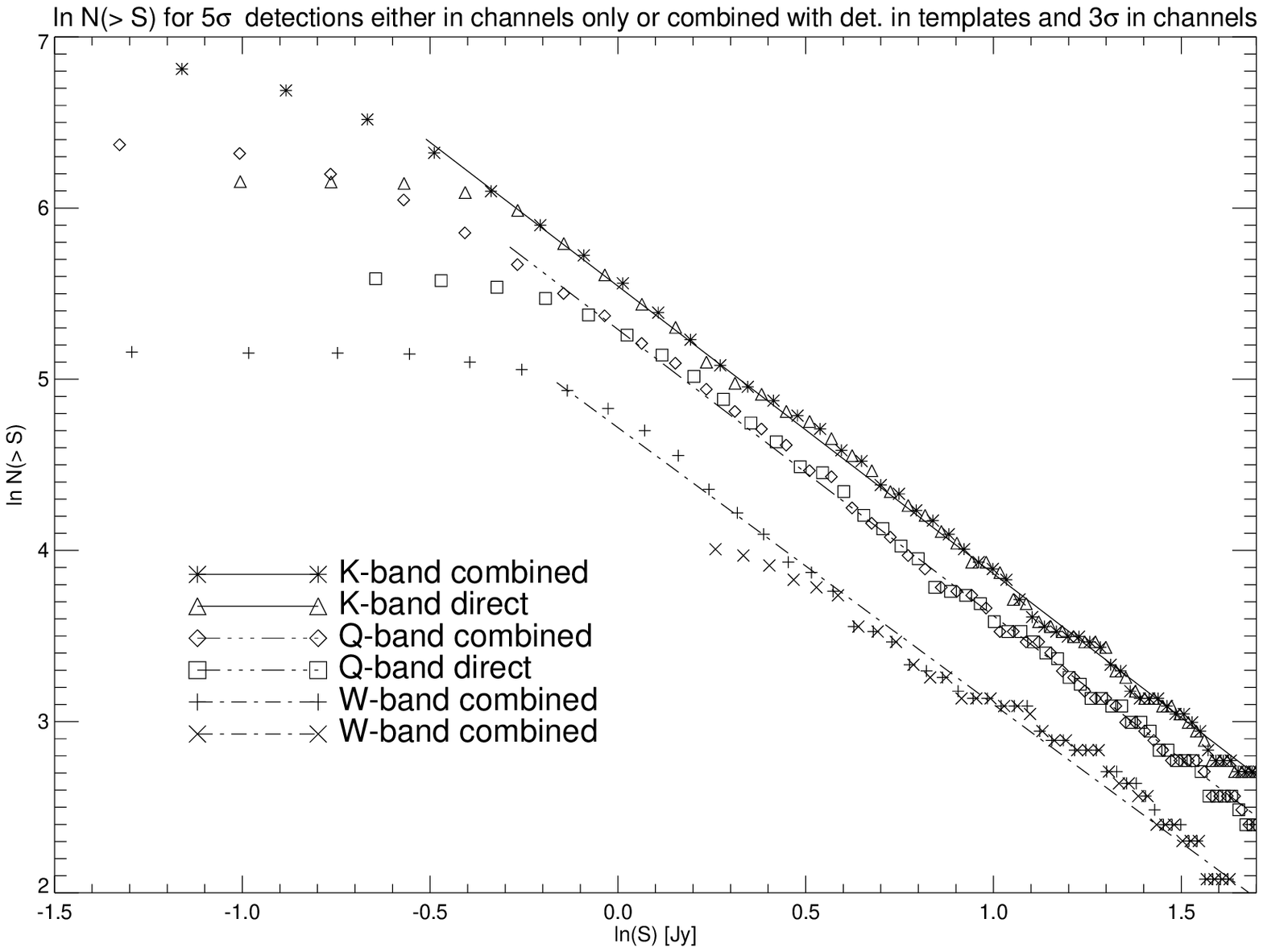}
\includegraphics[trim = 7mm 2mm 4mm 0mm, clip,scale=0.85]{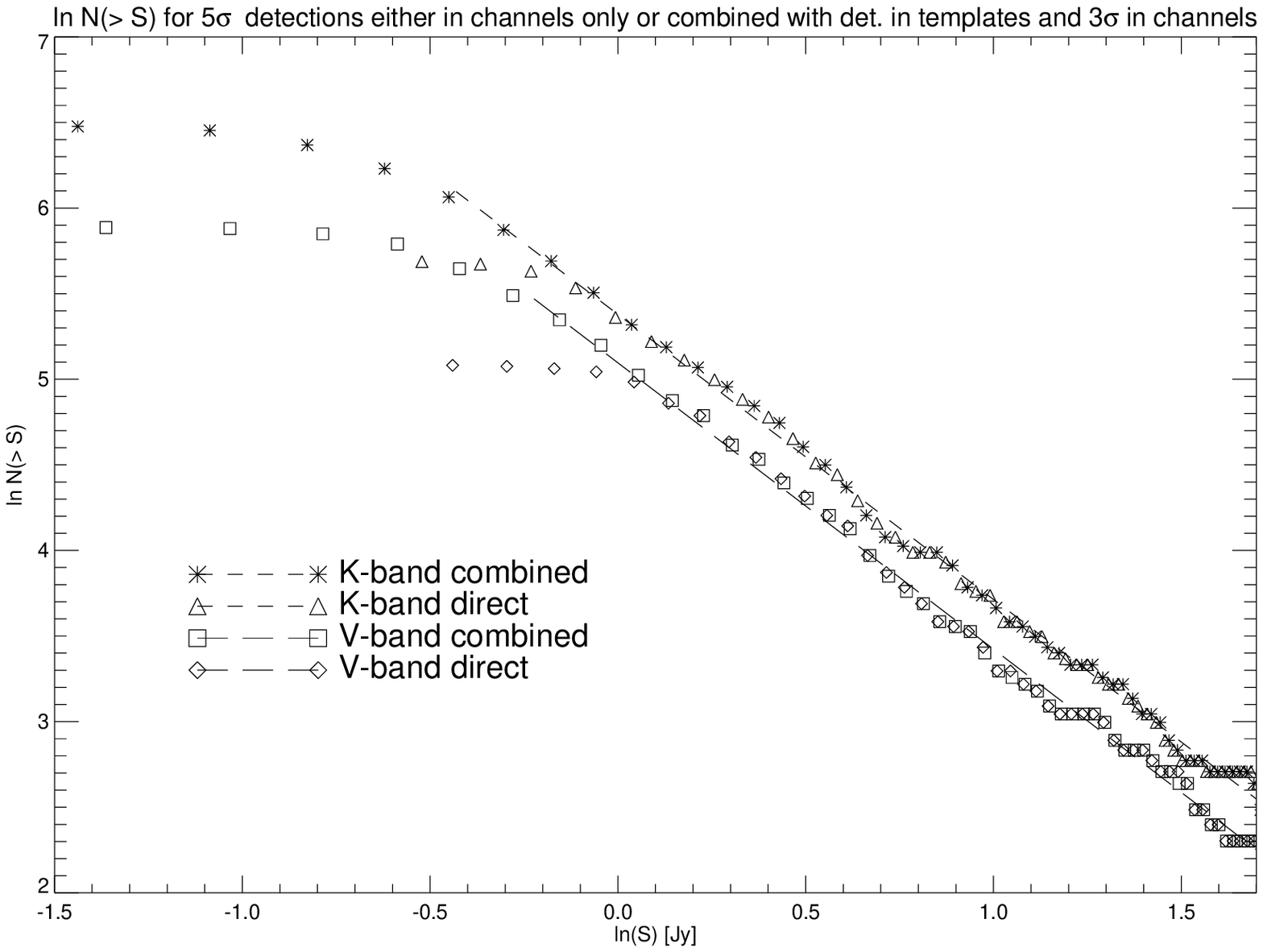}
\caption{Integral number counts of detected sources in WMAP 7 year data. ``Direct'' means only detections at $5\sigma$ in the channels; ``combined'' means detections either at $5\sigma$ in the channels or $5\sigma$ in the templates and $3\sigma$ in channels. The lines represent a linear fit to the number counts. Top: for K, Q and W band. Bottom: for Ka and V band. NB: counts are computed in bin-sizes of $\Delta S=0.1$ [Jy] and the axes are clipped in order to enhance the most interesting part. } 
\label{fig:complete}
\end{center}
\end{figure}

 In figures \ref{fig:complete} we highlight the advantage of combining the direct detections with detections in internal templates. Both plots provide the (logarithmic) integral counts for $5\sigma$ detections in the channels and for the combination of detections which are either $5\sigma$ in the channels or $5\sigma$ in the templates and $3\sigma$ in the channels. For visibility reasons, the top plot shows these integral counts for the K, Q and W band, while the bottom plot shows them for the Ka and V band. For the direct $5\sigma$ detections in the channels we see that our detections are complete approximately till {0.76,0.88,0.95,1.04,1.73} Jansky at respectively {K,Ka,Q,V,W} channels. For the combined integral counts we obtain complete sets of detections till approximately {0.60,0.64,0.72,0.80,0.84} Jansky in the respective channels. In both plots the lines represent a linear fit to the number counts, starting at the approximate completeness limit (of the combined counts).

\section{Conclusions}
\label{sec:concl}

We have developed a new procedure for detecting point sources in CMB data. First, we use needlet coefficients of the frequency maps optimized for point source detection and second, in addition to the frequency maps having CMB and noise as background we also use needlet coefficients of internal templates where the CMB is eliminated. In the frequency maps, only the sources with the strongest amplitude can be detected. In the internal templates however, the sources with the largest difference in amplitude from one frequency to another are more easily detected. In addition the pure instrumental noise background of the internal templates makes detection easier in many cases. In order to distinguish point sources from extended sources being parts of the diffuse galactic emission, we use $\chi^2$ tests to eliminate point like structures which do not have a beam like shape. In this paper we have applied this new point source detection procedure on the WMAP 7 year data.

We first used the beam and noise properties of the WMAP channels to optimize the kind of needlet to use for point source detection for each of the 5 WMAP frequency channels, as well as for the 5 internal template combinations with the highest signal-to-noise ratio. We then produced a set of simulated maps with simulated point sources, in order to test the detection procedure and to estimate error bars on amplitude and position.

We used a $5\sigma$ detection limit on the needlet coefficients in the frequency bands as well as in the internal templates. We detected in total 522 sources in the frequency bands only, and 2052 sources in the internal templates only, which all passed the rather conservative $\chi^2$ test. For sources which we detected at $5\sigma$ in the frequency maps, or at $5\sigma$ in the templates and at the same time $\geq3\sigma$ in the frequency maps, we estimated flux and position and attempted to identify them with sources found in other catalogues. All these 1116 sources are listed in table \ref{tab:new-PS} of which 1029 have a 5GHz or 1Jy counterpart. Of the remaining, 69 have only lower frequency-catalogues counterparts and 16 have no known lower-frequency counterpart. When comparing to catalogues based on WMAP data (the two catalogues obtained by the WMAP team (\cite{WMAPsrc}) and the NEWPS catalogues (\cite{mass})) or other observations at similar frequencies (the ERCSC based on Planck data (\cite{Planck})), we find 487 sources also found in the WMAP catalogues, 415 also found in the NEWPS\_5yr\_5s catalogue and 517 found also in the ERCSC (at at least one frequency of either 30,44,70,100 GHz). We point out that among the 49 sources in the WMAP catalogues which we do not detect/accept, only 13 are not detected, the remaining are either excluded by the $\chi^2$ test or the position is offset such that we are unable to identify it with our source. Finally, of the 1116 sources in table \ref{tab:new-PS} (also available at \verb=http:folk.uio.no/frodekh/PS_catalogue/Scodeller_PS_catalogue.txt=), 506 are not identified with sources in any of the catalogue based on WMAP or WMAP-like frequencies, but most are identified with a 5GHz counterpart. And 603 of 1116 sources have not been previously detected in WMAP data.

\section{Acknowledgement}
FKH acknowledges an OYI grant from the Norwegian Research Council; research by DM is supported by the European Research Council grant n.277742 (\emph{Pascal}). Super computers from NOTUR (The Norwegian metacenter for computational science) have been used in this work. We acknowledge the use of the HEALPix software package \cite{healpix} and the Legacy Archive for Microwave Background Data Analysis (LAMBDA) to retrieve the WMAP data set. This research has made use of data obtained from the High Energy Astrophysics Science Archive Research Center (HEASARC), provided by NASA's Goddard Space Flight Center and of the NASA/IPAC Extragalactic Database (NED) which is operated by the Jet Propulsion Laboratory, California Institute of Technology, under contract with the National Aeronautics and Space Administration.

\begin{deluxetable}{ccccccccllll}
\tabletypesize{\tiny}%scriptsize}%\footnotesize}
\tablecolumns{11}
\tablewidth{0pt}
\tablecaption{New PS's \label{tab:new-PS}}
\tablehead{   % column headings
 \colhead{\#} &
  \colhead{R.A [$^\circ$]}   &
  \colhead{Dec.[$^\circ$]} &
  %\colhead{$\varphi$}   &
  %\colhead{$\theta$}    &
  \colhead{K [Jy]}  &
  \colhead{Ka [Jy]}           &
  \colhead{Q [Jy]}  &
  \colhead{V [Jy]}        &
  \colhead{W [Jy]}    &
  %\colhead{Err K [Jy]}   &
  %\colhead{Err Ka [Jy]}    &
  %\colhead{Err Q [Jy]} &
  %\colhead{Err V [Jy]} &
  %\colhead{Err W [Jy]} &
  \colhead{F-flag} &
  \colhead{C-flag} &
  \colhead{Cat-ID} &
  \colhead{Off [']}
}
\startdata
   1  & 0.855    & -15.470  & \nodata                  & \nodata                  & \nodata                  & \nodata                  & 0.82$\pm$   0.25         & ....F  & ...  & PMN J0003-1517       & 11.1M.\\
   2  & 0.948    & -17.628  & \it{0.24}$\pm$\it{0.05}  & \nodata                  & 0.56$\pm$   0.16         & 0.66$\pm$   0.19         & \nodata                  & F.FF.  & ...  & PMN J0003-1727       & 12.1M.\\
   3  & 1.056    & -47.584  & 0.70$\pm$   0.16         & 0.89$\pm$   0.19         & 0.54$\pm$   0.14         & \nodata                  & \nodata                  & CCF..  & WN.  & PMN J0004-4736       & 4.0..\\
   4  & 1.079    & -11.820  & 0.38$\pm$   0.12         & 0.66$\pm$   0.17         & 0.59$\pm$   0.16         & \nodata                  & \nodata                  & FFF..  & ..P  & PMN J0004-1148       & 3.7..\\
   5  & 1.159    & 20.299   & \nodata                  & 0.43$\pm$   0.14         & \nodata                  & \nodata                  & \nodata                  & .F...  & ...  & GB6 J0004+2019       & 1.7..\\
   6  & 1.530    & -6.357   & 2.33$\pm$   0.16         & 2.47$\pm$   0.19         & 2.28$\pm$   0.18         & 1.96$\pm$   0.23         & 1.14$\pm$   0.25         & CCCCF  & WNP  & PMN J0006-0623       & 2.8M.\\
   7  & 1.578    & 38.319   & 0.50$\pm$   0.13         & 0.79$\pm$   0.17         & \nodata                  & \nodata                  & 1.07$\pm$   0.28         & FF..F  & ..P  & GB6 J0005+3820       & 4.3M.\\
   8  & 1.697    & -0.161   & 0.45$\pm$   0.13         & \nodata                  & \nodata                  & \nodata                  & \nodata                  & F....  & ...  & PMN J0006-0004       & 8.3..\\
\enddata
\end{deluxetable}

\end{document}